\RequirePackage{lineno}
\documentclass[prd,amsmath,amssymb,showpacs,superscriptaddress,nofootinbib,twocolumn]{revtex4-1}
\usepackage{graphicx}
\usepackage{epsfig}
\usepackage{dcolumn}
\usepackage{bm}
\usepackage{overpic}
\newcommand{\BR}{{\cal B}}

\newcommand{\LL}{\ell^+\ell^-}
\newcommand{\mystrut}{\rule[-.3\baselineskip]{0pt}{1.2\baselineskip}}

\usepackage[colorlinks,linkcolor=blue,anchorcolor=blue,citecolor=blue,dvipdfm]{hyperref} 

\begin{document}
\title{\boldmath Measurement of the $e^{+}e^{-} \to \eta J/\psi$ cross section and search for $e^{+}e^{-} \to \pi^{0} J/\psi$ at center-of-mass energies between 3.810 and 4.600~GeV}

\author{
  \begin{small}
    \begin{center}
      M.~Ablikim$^{1}$, M.~N.~Achasov$^{9,a}$, X.~C.~Ai$^{1}$,
      O.~Albayrak$^{5}$, M.~Albrecht$^{4}$, D.~J.~Ambrose$^{44}$,
      A.~Amoroso$^{48A,48C}$, F.~F.~An$^{1}$, Q.~An$^{45}$,
      J.~Z.~Bai$^{1}$
% ...
      \\
      \vspace{0.2cm}
      (BESIII Collaboration)\\
      \vspace{0.2cm} {\it
        $^{1}$ Institute of High Energy Physics, Beijing 100049, People's Republic of China\\
% ...
$^{4}$ Bochum Ruhr-University, D-44780 Bochum, Germany\\
        $^{5}$ Carnegie Mellon University, Pittsburgh, Pennsylvania 15213, USA\\
% ...
$^{9}$ G.I. Budker Institute of Nuclear Physics SB RAS (BINP), Novosibirsk 630090, Russia\\
% ...
$^{44}$ University of Rochester, Rochester, New York 14627, USA\\
        $^{45}$ University of Science and Technology of China, Hefei 230026, People's Republic of China\\
% ...
$^{48}$ (A)University of Turin, I-10125, Turin, Italy; (B)University of Eastern Piedmont, I-15121, Alessandria, Italy; (C)INFN, I-10125, Turin, Italy\\
% ...
$^{a}$ Also at the Novosibirsk State University, Novosibirsk, 630090, Russia\\
% ...
      }\end{center}
    \vspace{0.4cm}
  \end{small}
}

\affiliation{}

\date{\today}

\begin{abstract}
Using data samples collected with the BESIII detector operating at the BEPCII collider
at center-of-mass energies from 3.810 to 4.600 GeV, we perform a study of $e^{+}e^{-} \to \eta J/\psi$ and $\pi^0 J/\psi$.
Statistically significant signals of $e^{+}e^{-} \to \eta J/\psi$ are observed at $\sqrt{s}$ = 4.190, 4.210, 4.220, 4.230, 4.245, 4.260, 4.360 and 4.420 GeV,
while no signals of $e^{+}e^{-} \to \pi^{0} J/\psi$ are observed.
The measured energy-dependent Born cross section for $e^{+}e^{-} \to \eta J/\psi$ shows an enhancement around 4.2~GeV.
The measurement is compatible with an earlier measurement by Belle, but with a significantly improved precision.
\end{abstract}

\maketitle

\section{Introduction}
During the last decade, new charmoniumlike vector states, such as the
$Y(4260)$, $Y(4360)$ and $Y(4660)$, have been observed by BABAR~\cite{Y(4260)1,Y(4360)1},
Belle~\cite{Y(4260)2,Y(4260)21,Y(4360)2,Y(4630)2} and CLEO~\cite{Y(4260)3}.
The masses of these new $Y$ states are above the $D\bar D$ production threshold, ranging from 4.0 to 4.7 GeV/$c^{2}$.
Since all of them are produced in $e^{+}e^{-}$ annihilation (either directly or via the initial state radiation (ISR) process),
and since they have been observed to decay in dipion hadronic transitions to the $J/\psi$ or $\psi(3686)$,
one would naturally interpret these states as vector charmonium excitations.
However, peculiar features of these $Y$ states reveal an exotic nature that likely excludes a conventional charmonium interpretation.
These features include a discrepancy with the spectrum of vector charmonium states predicted by the potential model given in reference~\cite{speVect},
a surprisingly large coupling to final states without open-charm mesons~\cite{DDstate, hadfine},
and a lack of observation in the inclusive hadronic cross section~\cite{R-BESII}.
Also, very recently, several charged charmoniumlike states ---
the $Z_{c}$(3900)$^\pm$~\cite{Zc(3900)1,Y(4260)21,Zc(3900)12},
$Z_{c}$(3885)$^\pm$~\cite{Zc(3885)1},
$Z_{c}$(4020)$^\pm$~\cite{Zc(4020)1},
$Z_{c}$(4025)$^\pm$~\cite{Zc(4025)1}, as well as their isospin partners,
the neutral states $Z_c$(3900)$^0$~\cite{Zc(3900)12} and $Z_c$(4020)$^0$~\cite{Zc(4020)2} ---
were observed in the same mass region as these $Y$ states.
This suggests that the nature of the $Y$ states could be related to that of the $Z_{c}$ states.
Moreover, BESIII recently reported on the measurement of the cross sections of $e^{+}e^{-}\to\pi^+\pi^- h_c$~\cite{Zc(4020)1}
and $e^{+}e^{-}\to\omega\chi_{c0}$~\cite{OmegaChicj}.
The observed cross sections as a function of center-of-mass (CM) energy
are inconsistent with the line shape of $e^{+}e^{-}\to\pi^{+}\pi^{-}J/\psi$~\cite{Y(4260)1}.
These observations hint at the existence of a more complicated and mysterious underlying dynamics.

Many theoretical interpretations have been proposed to classify these $Y$ states, such as hybrid
charmonium~\cite{YNature1}, tetraquark~\cite{YNature5}, or hadronic molecule~\cite{YNature8} models,
but none of them has been able to describe all experimental observations in all aspects.
Searching for new decay modes and measuring the line shapes of their production cross sections will be very
helpful for these $Y$ states interpretation. Hadronic transitions (by $\eta$, $\pi^{0}$, or a pion pair) to lower charmonia like the $J/\psi$ are also regarded
as sensitive probes to study the properties of these $Y$ states~\cite{HT}.

The cross sections of $e^{+}e^{-} \to \eta J/\psi$ and $\pi^{0} J/\psi$ above the $D\bar D$ production
threshold have been evaluated within a non-relativistic framework derived from QCD~\cite{Theory1},
and their line shapes are predicted to be strongly affected by open charm effects~\cite{Theory2}.
Belle, BESIII and CLEO-c have measured the production cross sections of
$e^{+}e^{-} \to \eta J/\psi$ above the open charm threshold~\cite{belle,4040,etaJpsi1}.
However, Belle and CLEO-c results suffer from large statistical uncertainties.
BESIII reported on a more accurate result, but the measurement was limited to a single center of mass energy of $\sqrt{s}$ = 4.009~GeV.
Experimental studies with large data samples in a broad energy region may shed light on the nature of the $Y$ states.

In this paper, we report a measurement of the Born cross sections of
$e^{+}e^{-}\to\eta J/\psi$ and $\pi^{0} J/\psi$ from $e^{+}e^{-}$ CM energies 3.810 GeV to 4.600 GeV with data samples taken by BESIII.
In our analysis, the $\eta$ and $\pi^0$ are reconstructed in their two-photon decay mode and the $J/\psi$ via its decay into lepton pairs ($\ell^{+}\ell^{-}$).

\section{Detector and Monte Carlo simulation}
BEPCII~\cite{NIM:DET} is a double-ring $e^{+}e^{-}$ collider running at CM energies ranging from 2.0 to 4.6 GeV,
and providing a peak luminosity of $0.85\times10^{33}$ cm$^{-2}s^{-1}$ at the CM energy of $3.770$ GeV.
The BESIII~\cite{NIM:DET} detector has a geometrical acceptance of $93\%$ of $4\pi$ and has four main components.
(1) A small-cell, helium-based ($40\%$ He, $60\%$ C$_{3}$H$_{8}$) main drift chamber (MDC) with $43$ layers
provides an average single-hit resolution of $135$ $\mu$m, and a charged-particle momentum resolution in a $1$ T
magnetic field of $0.5\%$ at $1.0$ GeV$/c$. (2) A time-of-flight system (TOF) is constructed of $5$ cm thick plastic scintillators,
with $176$ detectors of $2.4$ m length in two layers in the barrel and $96$ fan-shaped detectors in the endcaps.
The barrel (endcap) time resolution of $80$ ps ($110$ ps) provides a $2\sigma$ $K/\pi$ separation for momenta up to $\sim$1 GeV$/c$.
(3) An electromagnetic calorimeter (EMC) consists of $6240$ CsI(Tl) crystals in a cylindrical structure (barrel) and two endcaps.
The photon energy resolution at $1.0$ GeV$/c$ is $2.5\%$ ($5\%$) in the barrel (endcaps),
and the position resolution is $6$ mm ($9$ mm) in the barrel (endcaps).
(4) The muon system (MUC) is located in the iron flux return yoke of the superconducting solenoid and consists of $1000$ m$^{2}$ of Resistive Plate Chambers (RPCs)
in nine barrel and eight endcap layers.  It provides a position resolution of $2.0$~cm.

The optimization of the selection criteria, the determination
of detection efficiencies, and the estimations of potential backgrounds are performed
based on Monte Carlo (MC) simulations taking various aspects of the experimental setup into account.
GEANT4-based~\cite{GEANT4} MC simulation software, which includes the geometric and material description of the
BESIII detector, the detector response and digitization models, as well as an accounting of the detector
running conditions and performances, is used to generate MC samples.
In the simulation, the electron-positron collision is simulated with the KKMC~\cite{KKMC,Evtgen} generator
taking into consideration the spread in the beam energy and ISR.
In this analysis, large signal MC samples of $e^{+}e^{-}\to\eta J/\psi$ and $e^{+}e^{-} \to \pi^0 J/\psi$
are generated at CM energies corresponding to the experimental values, where the line shape of the production cross section of these two
processes, assumed to be identical, are taken from the Belle experiment~\cite{belle}.

\section{Event Selection}
\label{seciii}
The candidate events of $e^{+}e^{-} \to \eta J/\psi$ and $\pi^{0} J/\psi$ are required to
have two charged tracks with a total net charge of zero and at least two photon candidates.

Charged tracks are reconstructed from the hits in the MDC. Each charged track is required to
have a polar angle that is well within the fiducial volume of the MDC, $|\cos\theta|<0.93$,
where $\theta$ is the polar angle of the track in the laboratory frame,
to have a point of closest approach to the interaction point that is within
$\pm10$ cm along the beam direction and within $1$ cm
in the radial direction, and to have a momentum $p$ larger than 1.0 GeV/$c$.
Electron and muon separation is carried out by making use of the deposited energy in the EMC.
Tracks with an energy deposition of $E<0.4$ GeV are identified as muons,
while tracks with $E/p>0.8~c$ are identified as electrons or positrons.

Photon candidates are reconstructed by isolated showers in the EMC.
The photon energy is required to be at least 25 MeV in the barrel ($|\cos\theta|<0.80$)
and 50 MeV in the endcaps ($0.86<|\cos\theta|<0.92$).
To eliminate showers produced by charged particles, the angle between the shower and
the nearest charged track must be larger than 20 degrees.
To suppress electronic noise and energy depositions unrelated to the physical event,
the EMC time $t$ of the photon candidate must be in coincidence with collision events,
in the range from 0 $\leq$ $t$ $\leq$ 700 ns.

A kinematic fit that imposes momentum and energy conservation (4C) is implemented under the
hypothesis of $e^{+}e^{-} \to \gamma \gamma \ell^+\ell^-$ to improve the momentum and
energy resolutions of the final-state particles and to reduce the potential backgrounds.
The chi-square of the kinematic fit, $\chi^{2}_{4C}$, is required to be less than $40$.
If there are more than two photons in an event, the combination of $\gamma \gamma \ell^{+}\ell^{-}$
with the least $\chi^{2}_{4C}$ is chosen.
To suppress the backgrounds from radiative Bhabha and radiative dimuon events associated
with a random photon candidate, the energy of each selected photon is further required to be larger than 80 MeV.

\begin{figure*}[htbp]
\begin{center}
\begin{overpic}[width=7.0cm,height=5.0cm,angle=0]{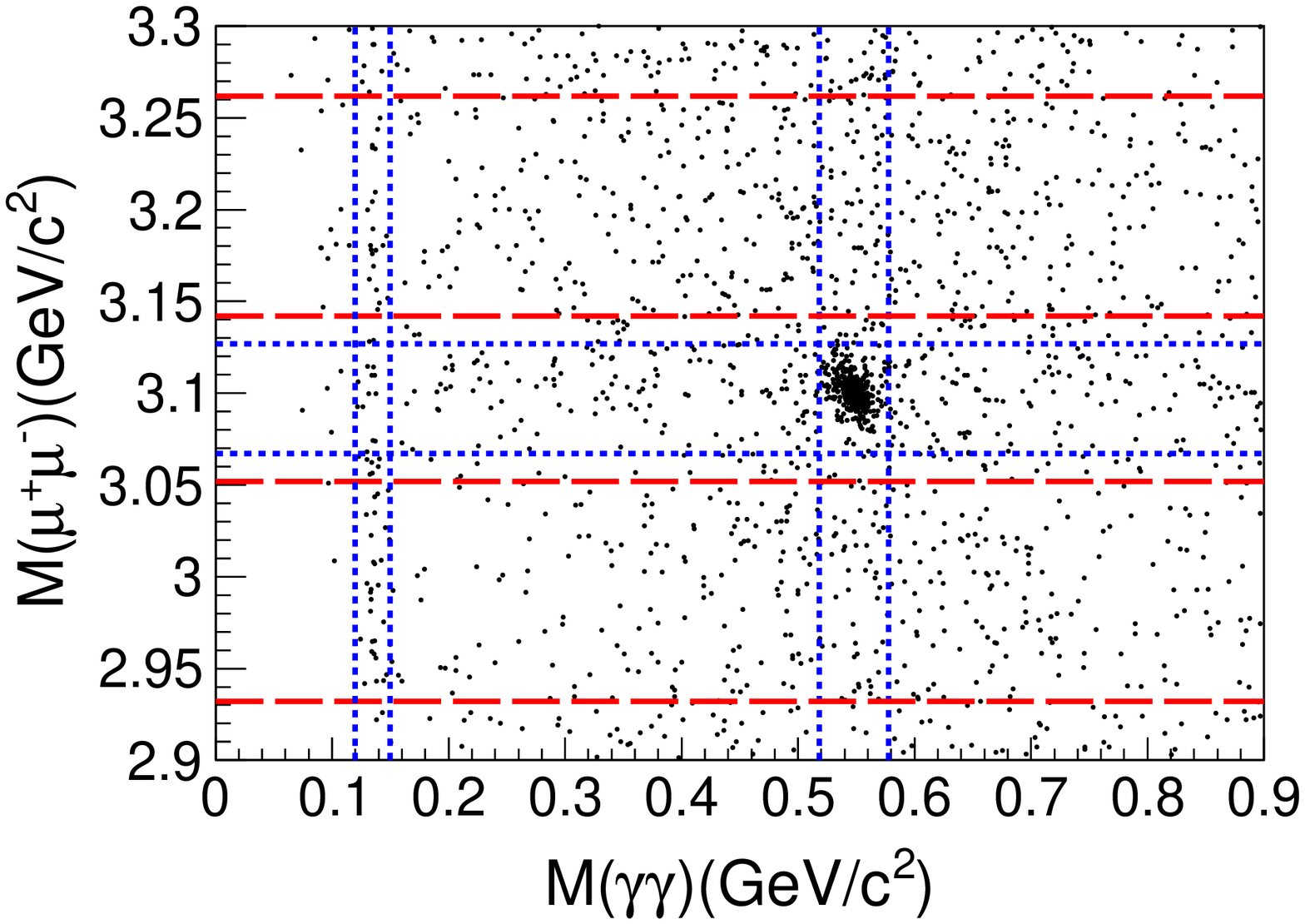}
\put(19,57){\large\bf (a)}
\end{overpic}
\begin{overpic}[width=7.0cm,height=5.0cm,angle=0]{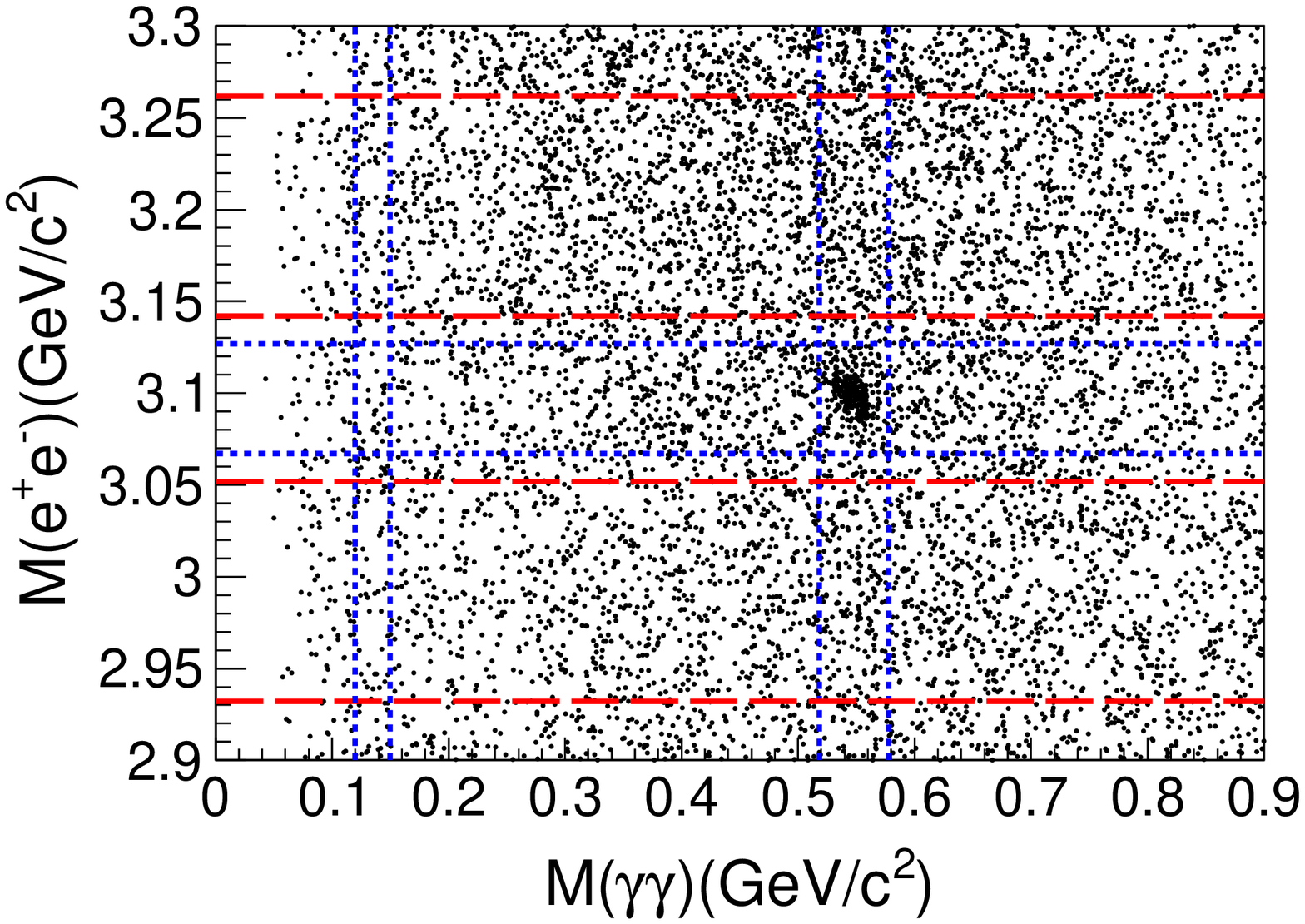}
\put(19,57){\large\bf (b)}
\end{overpic}
\begin{overpic}[width=7.0cm,height=5.0cm,angle=0]{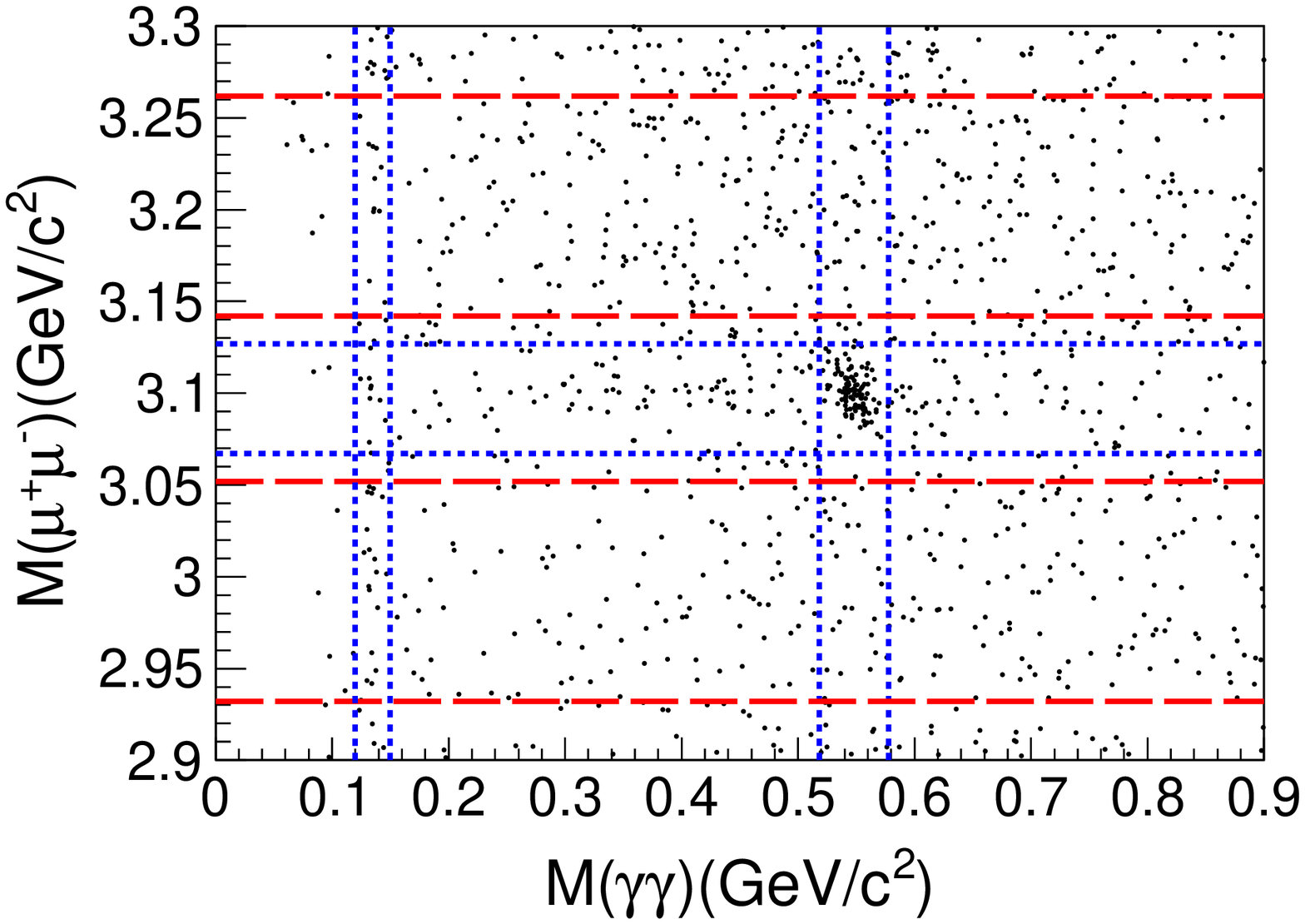}
\put(19,57){\large\bf (c)}
\end{overpic}
\begin{overpic}[width=7.0cm,height=5.0cm,angle=0]{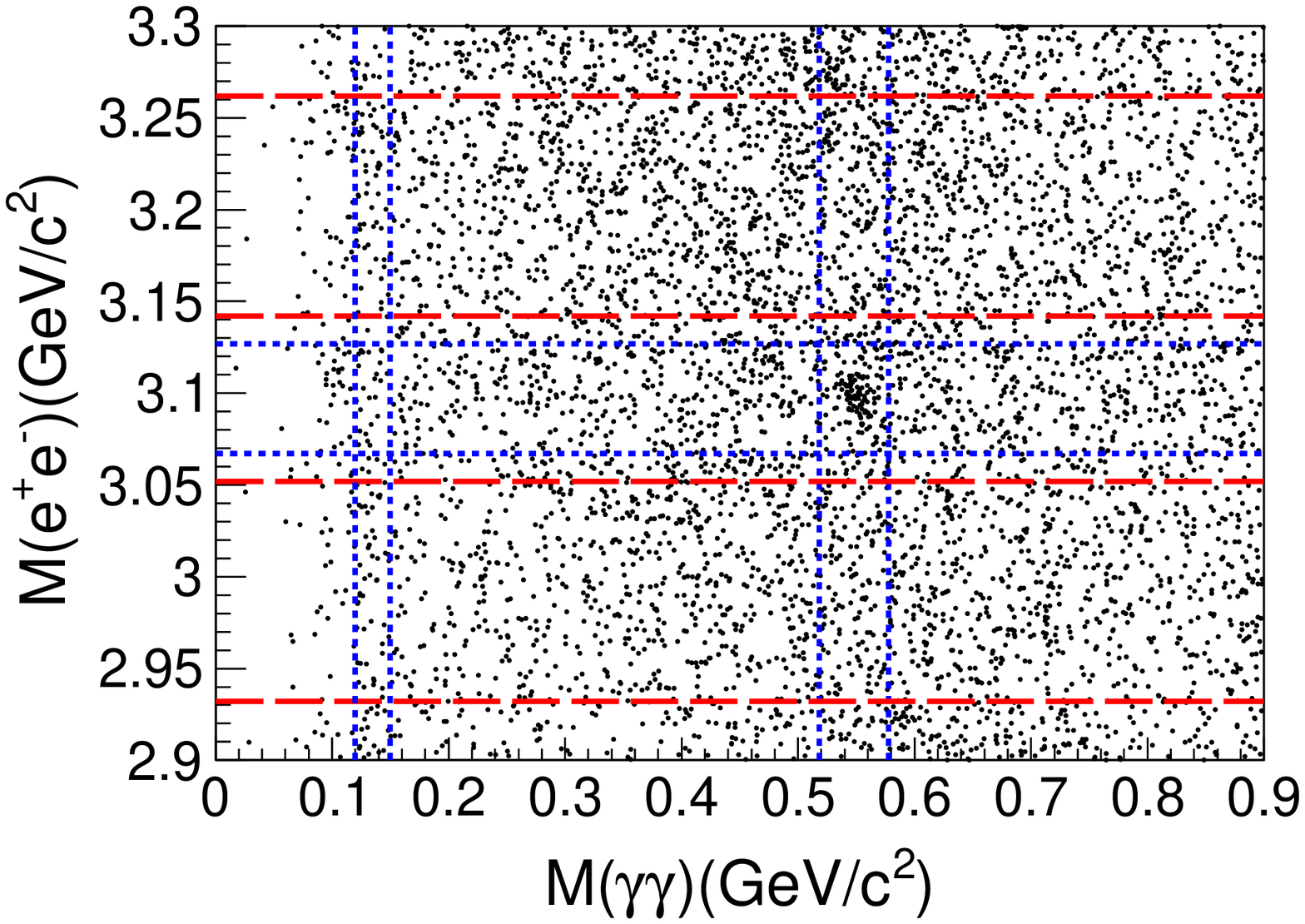}
\put(19,57){\large\bf (d)}
\end{overpic}
\end{center}
\caption{Scatter plots of $M(\ell^{+}\ell^{-})$ versus $M(\gamma\gamma)$ for data at $\sqrt{s}$ = 4.230 (top panels (a, b)) and 4.260 GeV (bottom panels (c, d)). The two panels on the left-hand side correspond to the $\mu^{+}\mu^{-}$ mode and the right-hand side to the $e^{+}e^{-}$ mode.
The blue dotted lines denote the $\eta/\pi^{0}$ and $J/\psi$ mass bands. The red dashed lines denote the sideband regions of $J/\psi$.}\label{scatter}
\end{figure*}

Figure~\ref{scatter} depicts scatter plots of the invariant mass of lepton pairs, $M(\ell^+\ell^-)$,
versus that of two photons, $M(\gamma\gamma)$, for data taken at $\sqrt{s}$ = 4.230 and 4.260 GeV.
A clear accumulation of events is observed around the intersection of the $\eta$ and $J/\psi$ mass regions,
which indicates $e^+e^-\to\eta J/\psi$ signals.
There is no significant signal observed around the intersection of the $\pi^0$ and $J/\psi$ mass regions.
MC studies show that dominant backgrounds are from the radiative Bhabha and dimuon events,
and are expected to be distributed uniformly around the $J/\psi$ and $\eta/\pi^{0}$ mass regions.
A significantly larger background yield is observed in the $e^{+}e^{-}$ mode than in the $\mu^{+}\mu^{-}$ mode,
which is due to the much larger Bhabha scattering cross section compared with the dimuon cross section.
MC simulations show that the resolution of the invariant mass distributions of lepton pairs
is about 10.7 MeV/c$^{2}$ for the $\mu^{+}\mu^{-}$ mode and 11.5 MeV/c$^{2}$ for the $e^{+}e^{-}$ mode.
The candidate event of $e^+e^-\to\eta J/\psi$ is required to be within the $J/\psi$ signal region,
defined as $3.067<M(\ell^{+}\ell^{-})<3.127$ GeV/$c^{2}$.
Sideband regions, defined as $2.932<M(\ell^{+}\ell^{-})<3.052$ GeV/$c^{2}$
and $3.142<M(\ell^{+}\ell^{-})<3.262$ GeV/$c^{2}$, four times as wide as the signal region,
are used to estimate the non-$J/\psi$ background contributions.

\begin{figure*}[htbp]
\begin{center}
\begin{overpic}[width=7.0cm,height=5.0cm,angle=0]{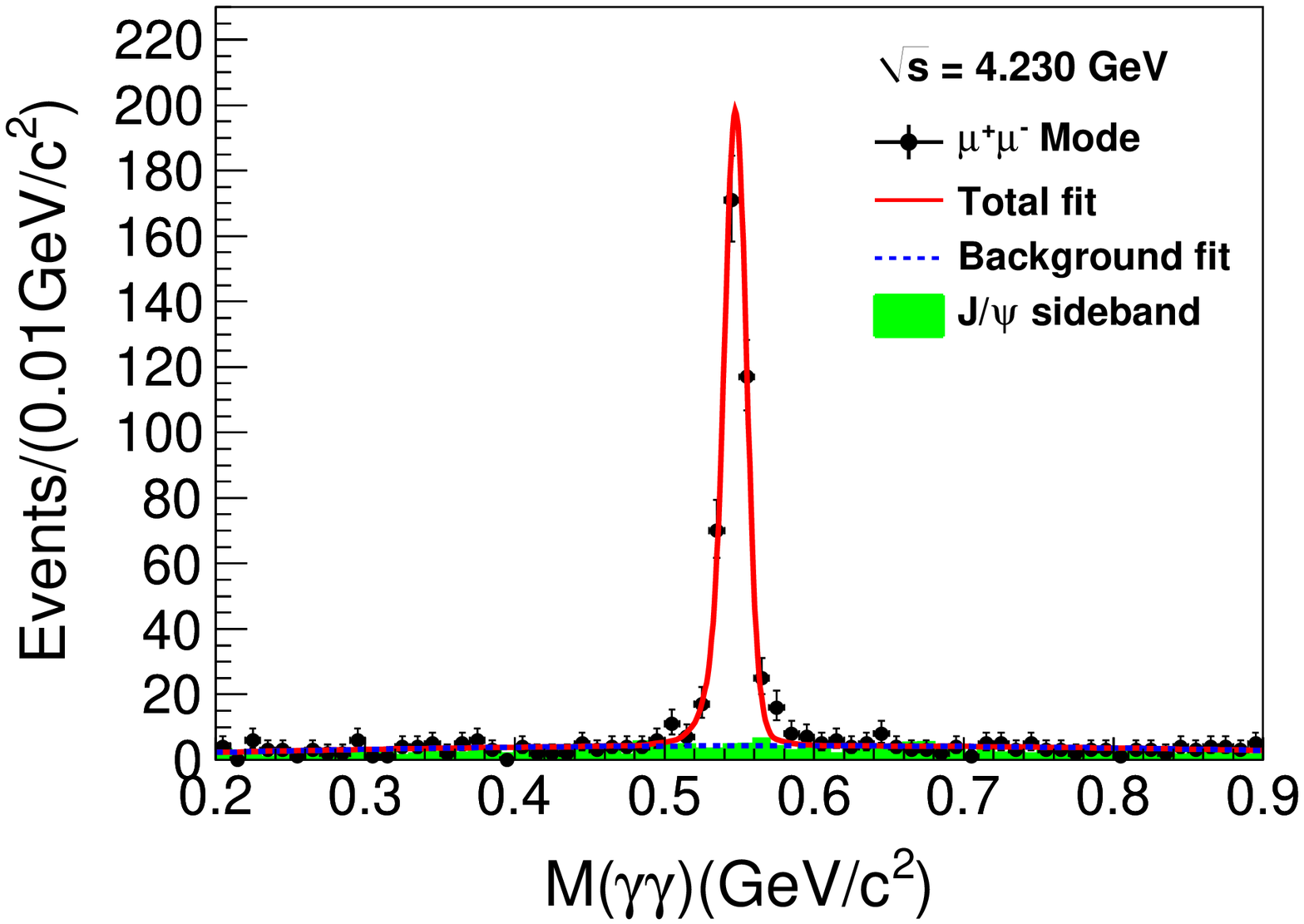}
\put(19,57){\large\bf (a)}
\end{overpic}
\begin{overpic}[width=7.0cm,height=5.0cm,angle=0]{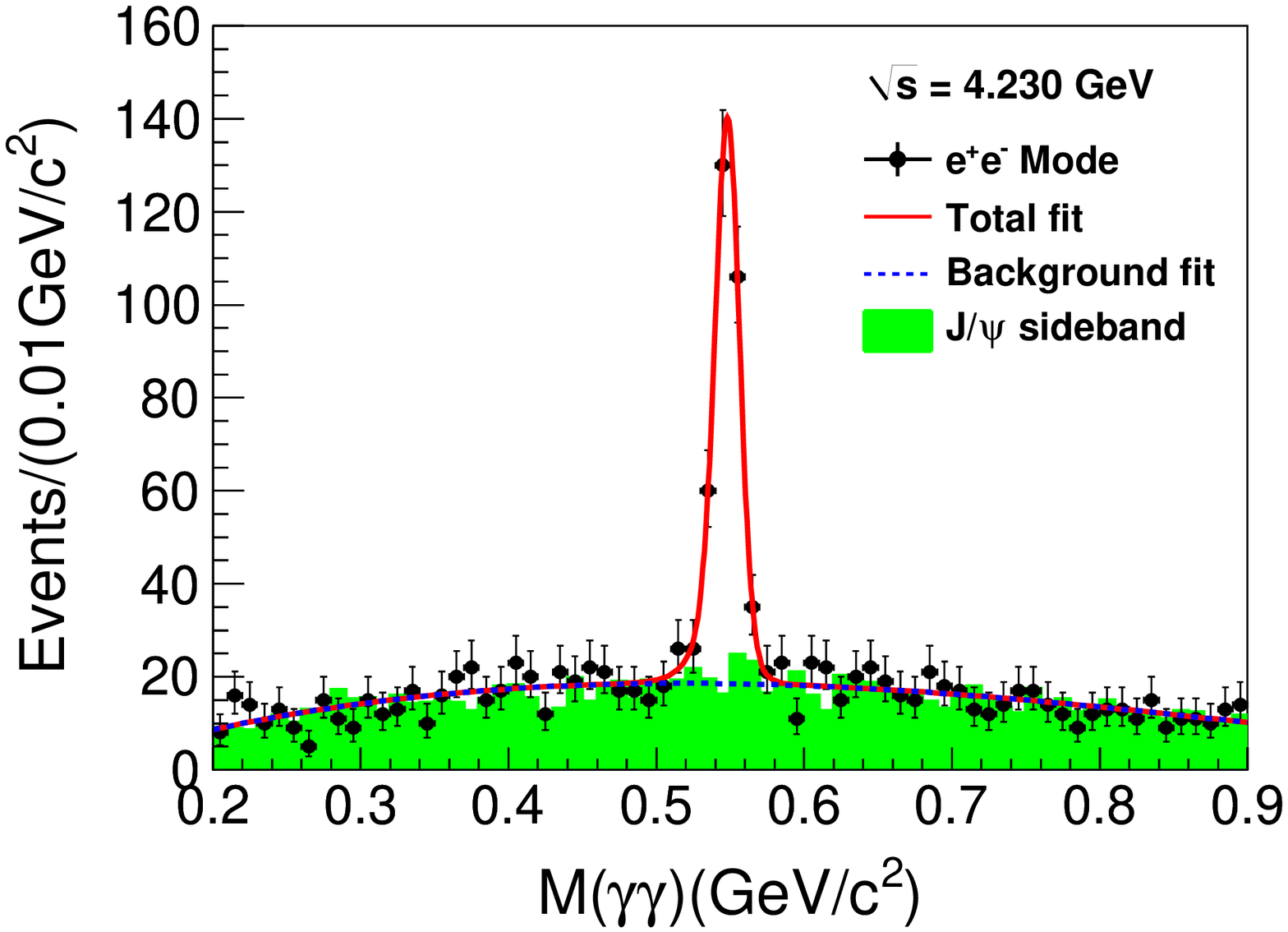}
\put(19,57){\large\bf (b)}
\end{overpic}
\begin{overpic}[width=7.0cm,height=5.0cm,angle=0]{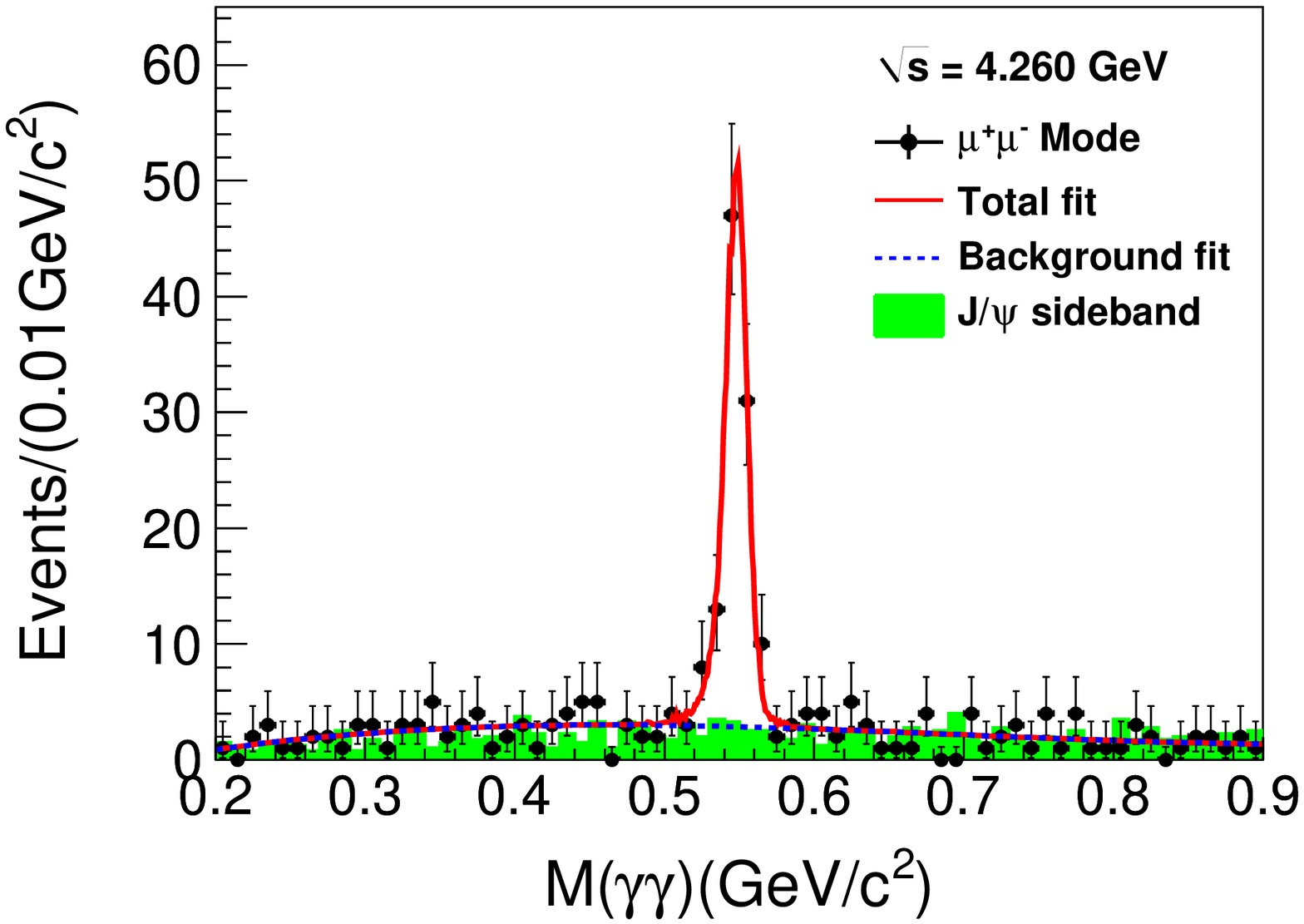}
\put(19,57){\large\bf (c)}
\end{overpic}
\begin{overpic}[width=7.0cm,height=5.0cm,angle=0]{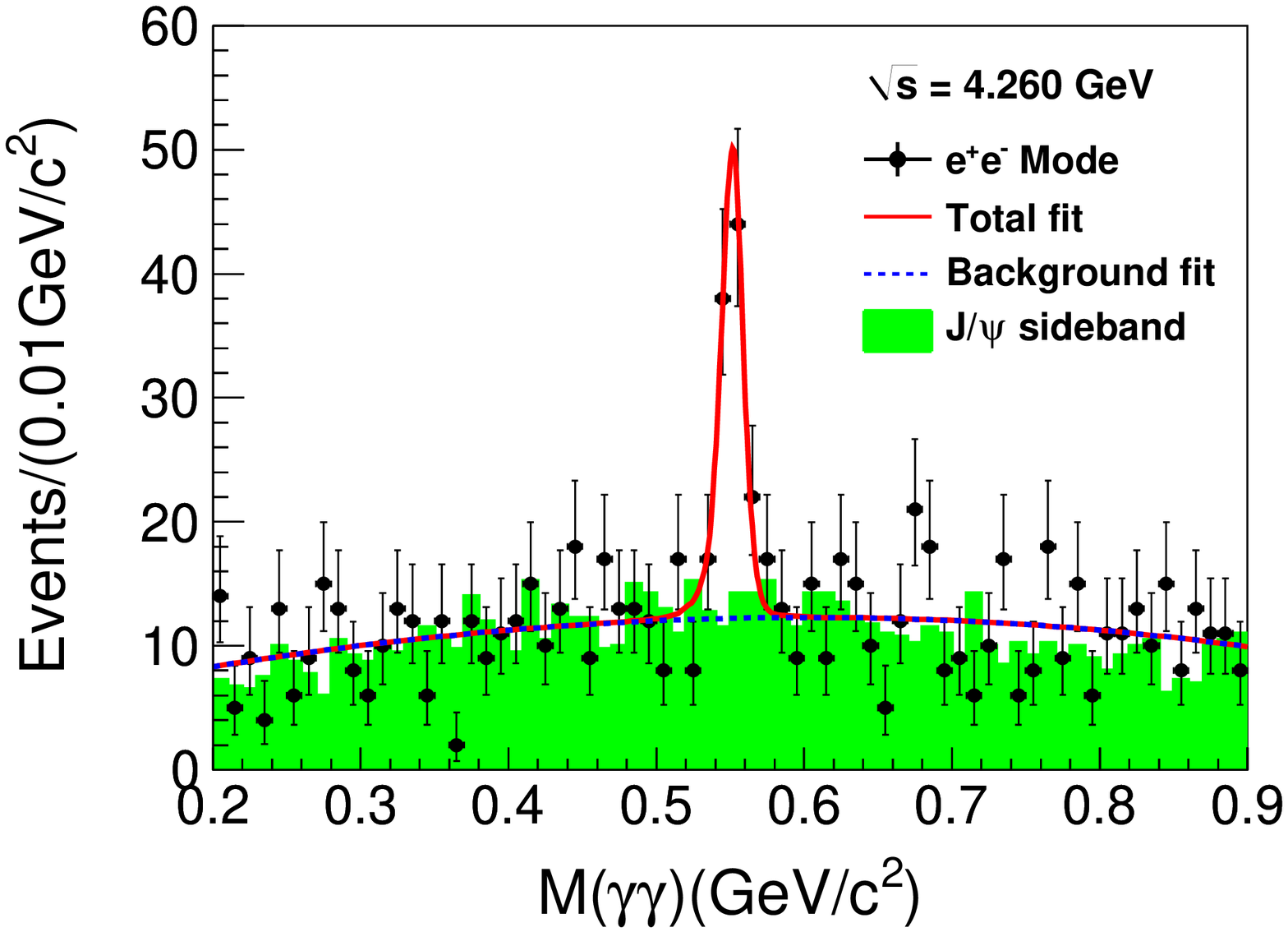}
\put(19,57){\large\bf (d)}
\end{overpic}
\end{center}
\caption{Invariant mass distributions of two photons for data at $\sqrt{s}$ = 4.230 (top panels (a, b)) and 4.260 GeV (bottom panels (c, d)). The left two plots are for the $J/\psi\to\mu^{+}\mu^{-}$ mode and the right two for $J/\psi\to e^{+}e^{-}$ mode.
Dots with error bars are for data in the $J/\psi$ signal region, the green shaded histograms for the normalized $J/\psi$ sideband events, the red solid curves for the total fit results and the blue dotted curves for the background from the fit.}
\label{4230gg}
\end{figure*}

After selecting the $J/\psi$ signal, the invariant mass distributions of two photons,
$M(\gamma \gamma)$, are shown in Fig.~\ref{4230gg} for data at $\sqrt{s}$ = 4.230 and 4.260 GeV.
Clear $\eta$ signals are observed.
The corresponding normalized distributions from the events in the $J/\psi$ sideband regions are shown as shaded histograms in the plots.
The backgrounds are well described by $J/\psi$ sideband events and show no peaking structure within the $\eta$ signal region.

\begin{figure*}[htbp]
\begin{center}
\begin{overpic}[width=7.0cm,height=5.0cm,angle=0]{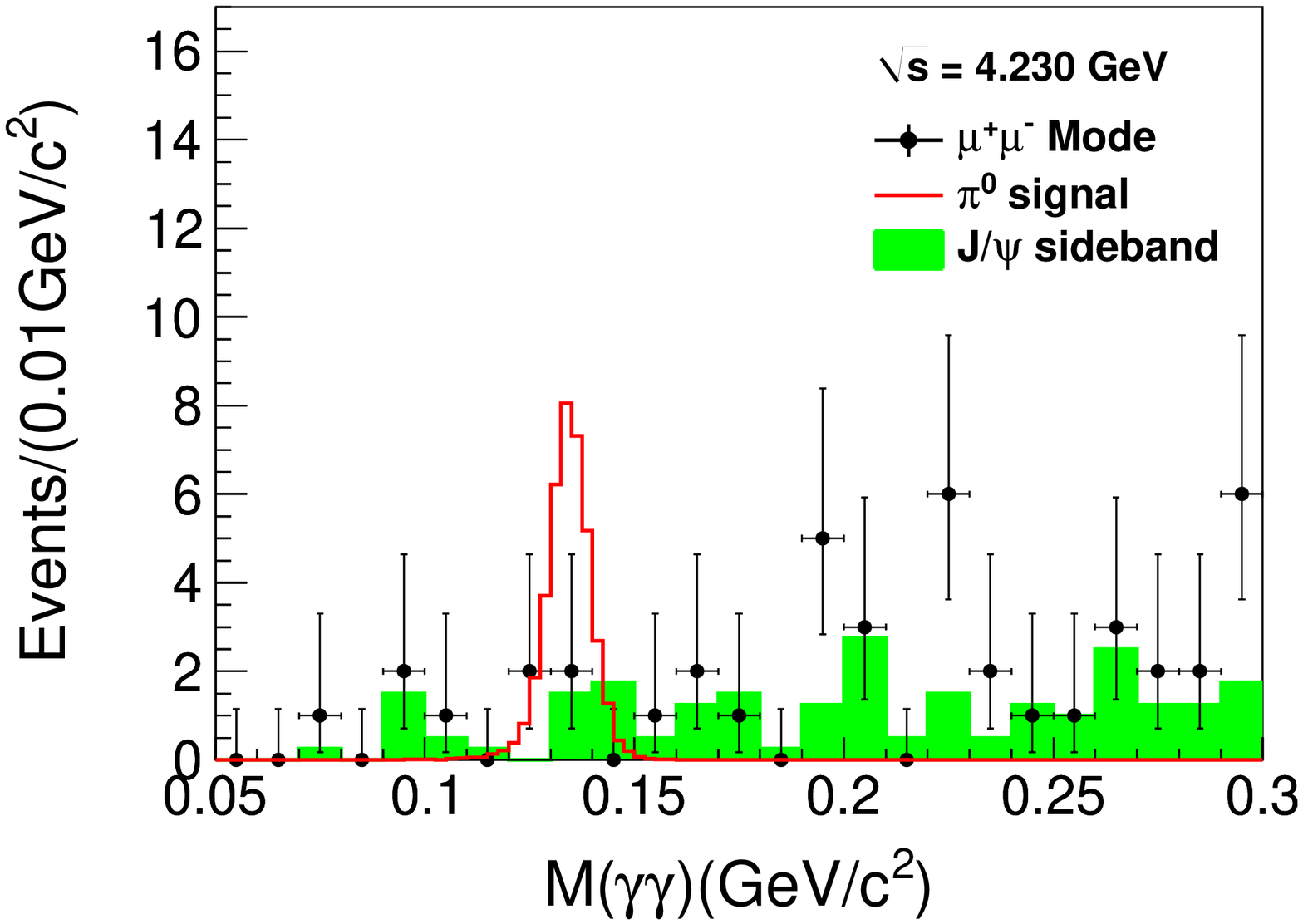}
\put(19,57){\large\bf (a)}
\end{overpic}
\begin{overpic}[width=7.0cm,height=5.0cm,angle=0]{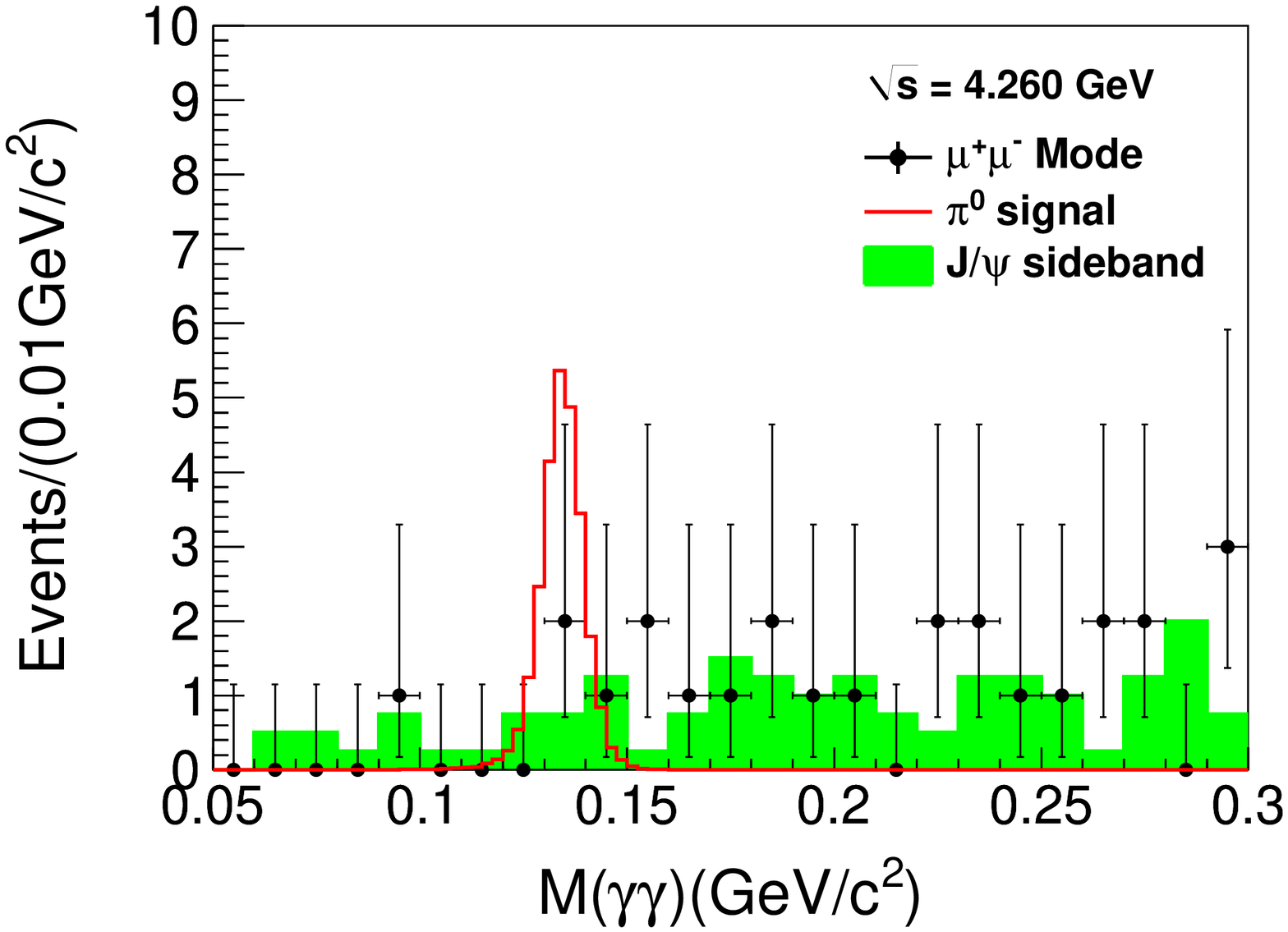}
\put(19,57){\large\bf (b)}
\end{overpic}
\end{center}
\caption{Invariant mass distributions of two photons for data at $\sqrt{s}$ = 4.230 (a) and 4.260 GeV (b) in $J/\psi\to\mu^{+}\mu^{-}$ mode. Dots with error
  bars are for data in the $J/\psi$ signal region, the green shaded histograms are the normalized $J/\psi$ sideband events and the red histograms are $\pi^{0}$
  MC signal with arbitrary normalization.}\label{4230ggPi0}
\end{figure*}

The process $e^+e^-\to\pi^0 J/\psi$ is also searched for in the $J/\psi\to\mu^+\mu^-$
mode by analyzing the $M(\gamma \gamma)$ distribution around the $\pi^0$ mass region.
Such a search is not performed for the $J/\psi\to e^+e^-$ mode due to the large background of radiative Bhabha events.
Due to the misidentification of $\pi^{\pm}$ as $\mu^{\pm}$, peaking background from
$e^{+}e^{-} \to \pi^{+}\pi^{-}\pi^{0}$ would contaminate the $\pi^0$ signal for
both candidate events within the $J/\psi$ signal or sideband regions.
To remove such backgrounds, we require that at least one charged track has a muon counter hit depth larger than $30$ cm.
Figure~\ref{4230ggPi0} shows the $M(\gamma \gamma)$ distributions around the $\pi^0$ mass
region after this requirement. No significant signal is observed for $\pi^{0}\to\gamma\gamma$ decays.

\section{\boldmath Fits to the M($\gamma \gamma$) spectrum and cross section results}
After imposing the $J/\psi$ signal selection, an unbinned maximum likelihood fit is performed on $M(\gamma\gamma)$ in the $J/\psi\to e^{+}e^{-}$
and $\mu^{+}\mu^{-}$ modes, respectively.
The probability density function (PDF) of the $M(\gamma\gamma)$ distribution for $\eta$ signals
is obtained from signal MC simulations convoluted with a Gaussian function, where the Gaussian function describes
the difference in resolution between data and MC simulation, and its parameters are left free in the fit.
The background shape is described by a second-order Chebyshev polynomial function.
The corresponding fit results for $\sqrt{s}$ = 4.230 and 4.260 GeV are shown in Fig.~\ref{4230gg}
and the numbers of $\eta$ signal events are summarized in Table~\ref{ResultsEta}.
The statistical significances of $\eta$ signals are larger than $8\sigma$,
which are examined using the differences in log-likelihood values of fits with or without an $\eta$ signal component.

\begin{table*}[htbp]
  \footnotesize
  \centering
  \caption{Results on $e^{+}e^{-}\to\eta J/\psi$ in data samples in which a signal is observed with a statistical significance larger than $5\sigma$. The table shows the CM energy $\sqrt{s}$, integrated luminosity $\mathcal{L}_\mathrm{int}$, number of observed $\eta$ events $N^\mathrm{obs}_{\eta}(\mu^{+}\mu^{-})$/$N^\mathrm{obs}_{\eta}(e^{+}e^{-})$ from the fit, efficiency $\epsilon_{\mu}/\epsilon_{e}$, radiative correction factor $(1+\delta^{r})$, vacuum polarization factor $(1+\delta^{v})$, Born cross section $\sigma^{B}(\mu^{+}\mu^{-})$/$\sigma^{B}(e^{+}e^{-})$  and combined Born cross section $\sigma^{B}_\mathrm{Com}$. The first uncertainties are statistical and the second systematic.}
  \begin{tabular}{ccccccccccc}
  \hline
  \hline
  \mystrut
  $\sqrt{s}$(GeV) &$\mathcal{L}$(pb$^{-1})$ &$N^\mathrm{obs}_{\eta}(\mu^{+}\mu^{-})$ &$N^\mathrm{obs}_{\eta}(e^{+}e^{-})$ &$\epsilon_{\mu}(\%)$&$\epsilon_{e}(\%)$ &$(1+\delta^{r})$ &$(1+\delta^{v})$ &$\sigma^{B}(\mu^{+}\mu^{-})$(pb) &$\sigma^{B}(e^{+}e^{-})$(pb)  &$\sigma^{B}_\mathrm{Com}$(pb)\\
  \hline
  $4.190$    &$43.1$    &$17.5\pm4.3$ &$10.4\pm3.6$     &$35.2$ &$24.1$      &$0.866$   &$1.056$   &$53.7\pm13.2\pm3.1$     &$46.6\pm16.1\pm1.7$       &$50.8\pm10.2\pm2.1$\\
  $4.210$    &$54.6$    &$25.7\pm5.1$ &$14.8\pm4.5$     &$33.7$ &$23.1$      &$0.914$   &$1.057$   &$61.6\pm12.2\pm4.1$     &$51.7\pm15.7\pm4.5$       &$57.8\pm9.6\pm3.2$\\
  $4.220$    &$54.1$    &$32.6\pm5.8$ &$11.4\pm3.9$     &$33.1$ &$22.8$      &$0.937$   &$1.057$   &$78.2\pm13.9\pm5.0$     &$39.6\pm13.6\pm2.9$       &$57.7\pm9.7\pm3.0$\\
  $4.230$    &$1091.7$  &$394.3\pm20.9$ &$274.9\pm20.1$ &$32.4$ &$22.3$      &$0.960$   &$1.056$   &$46.8\pm2.5\pm2.5$      &$47.3\pm3.5\pm3.4$        &$47.0\pm2.0\pm2.2$\\
  $4.245$    &$55.6$    &$9.3\pm3.3$ &$9.7\pm3.6$       &$31.4$ &$21.7$      &$0.992$   &$1.056$   &$21.6\pm7.7\pm2.4$      &$32.6\pm12.1\pm3.5$       &$24.8\pm6.5\pm2.0$\\
  $4.260$    &$825.7$   &$94.4\pm10.5$ &$75.9\pm11.9$   &$30.3$ &$20.9$      &$1.021$   &$1.054$   &$14.9\pm1.7\pm1.1$      &$17.4\pm2.7\pm1.2$        &$15.7\pm1.4\pm0.9$\\
  $4.360$    &$539.8$   &$19.8\pm5.3$ &$23.9\pm7.7$     &$25.7$ &$17.7$      &$1.168$   &$1.051$   &$4.9\pm1.3\pm0.7$       &$8.7\pm2.8\pm1.1$         &$5.6\pm1.2\pm0.6$\\
  $4.420$    &$1074.7$  &$56.9\pm8.2$ &$42.6\pm9.9$     &$24.2$ &$16.7$      &$1.225$   &$1.053$   &$7.3\pm1.1\pm0.8$       &$7.8\pm1.8\pm0.7$         &$7.5\pm0.9\pm0.6$\\
  \hline
  \hline
    \end{tabular}
    \label{ResultsEta}
\end{table*}

The same event selection criteria are implemented on the other 15 data samples taken at different CM energies.
We observe a significant yield of $\eta$ signal with more than $5\sigma$ statistical significance for data
at $\sqrt{s}$ = 4.190, 4.210, 4.220, 4.245, 4.360 and 4.420 GeV. Using the same fit procedure,
the numbers of $\eta$ signal events for these energies are also obtained and listed in Table~\ref{ResultsEta}.
There are no significant $\eta$ signals observed for the other 9 energy points
($\sqrt{s}$ = 3.810, 3.900, 4.090, 4.310, 4.390, 4.470, 4.530, 4.575, 4.600 GeV), and the upper limits at the
90\% confidence level (C.L.) on the Born cross section are determined with the $J/\psi\to\mu^+\mu^-$ decay mode only,
due to the large background from Bhabha events in $J/\psi \to e^+e^-$.
Since the statistics are low for the no-signal data samples, the number of observed events is obtained by
counting the entries in the $\eta$ signal region ($0.518<M(\gamma\gamma)<0.578$ GeV/c$^2$).
The number of background events in the signal region is estimated
by the events in the $\eta$ sideband region or $J/\psi$ sideband region (with an additional $\eta$
signal mass window requirement) by assuming a flat distribution of background around signal regions.
The $\eta$ sideband region is defined as $0.383<M(\gamma\gamma)<0.503$ GeV/c$^2$ and
$0.593<M(\gamma\gamma)<0.713$ GeV/c$^2$, where their sizes are four times as that of the signal region.
The results are all listed in Table~\ref{ResultsEtaUpper}.

\begin{table*}[htbp]
  \centering
  \caption{Upper limits of $e^{+}e^{-} \to \eta J/\psi$ using the $\mu^{+}\mu^{-}$ mode. The table shows the CM energy $\sqrt{s}$, integrated luminosity $\mathcal{L}_\mathrm{int}$, number of observed $\eta$ events $N^\mathrm{sg}_{\eta}$, number of background from $\eta$ sideband $N^\mathrm{sb}_{\eta}$, and from $J/\psi$ sideband $N^\mathrm{sb}_{J/\psi}$, efficiency $\epsilon$, upper limit of signal number with the consideration of selection efficiency $N^\mathrm{up}_{\eta}/\epsilon$ (at the $90\%$ C.L.), radiative correction factor $(1+\delta^{r})$, vacuum polarization factor $(1+\delta^{v})$, Born cross section $\sigma^{B}$ and upper limit on the Born cross sections $\sigma^{B}_\mathrm{up}$ (at the $90\%$ C.L.). The first uncertainties are statistical and the second systematic.}
  \begin{tabular}{ccccccccccc}
  \hline
  \hline
  \mystrut
  ~$\sqrt{s}$(GeV)~ &~$\mathcal{L}(pb^{-1})$~&~$N^\mathrm{sg}_{\eta}$~ &~$N^\mathrm{sb}_{\eta}$~ &~$N^\mathrm{sb}_{J/\psi}$~  &~$\epsilon (\%)$~&~~~$N^\mathrm{up}_{\eta}/\epsilon$~&~$(1+\delta^{r})$~ &~$(1+\delta^{v})$~ &~$\sigma^{B}$(pb)~ &~$\sigma^{B}_\mathrm{up}$(pb)~\\
  \hline
  $3.810$     &$50.5$      &$5$         &$9$    &$11$   &$32.3$   &$<23.3$        &$1.243$       &$1.056$       &$5.5^{+7.1}_{-4.6}\pm0.2$                     &$<15.1$\\
  $3.900$     &$52.6$      &$5$         &$8$    &$7$    &$38.3$   &$<20.9$        &$0.775$       &$1.049$       &$7.9^{+9.2}_{-5.9}\pm0.3$                     &$<20.8$\\
  $4.090$     &$52.6$      &$7$         &$7$    &$5$    &$31.0$   &$<36.3$        &$1.087$       &$1.052$       &$12.2^{+9.0}_{-6.2}\pm0.9$                    &$<25.9$\\
  $4.310$     &$44.9$      &$1$         &$4$    &$2$    &$27.4$   &$<11.7$        &$1.105$       &$1.053$       &$0.0^{+7.2}_{-2.9}\pm0.0$                     &$<9.5$\\
  $4.390$     &$55.2$      &$5$         &$1$    &$4$    &$25.0$   &$<38.4$        &$1.198$       &$1.051$       &$11.7^{+8.5}_{-5.4}\pm0.6$                    &$<23.5$\\
  $4.470$     &$109.9$     &$2$         &$12$   &$8$    &$23.5$   &$<14.7$        &$1.258$       &$1.055$       &$-1.2^{+3.5}_{-1.9}\pm0.1$                    &$<4.3$\\
  $4.530$     &$110.0$     &$5$         &$6$    &$4$    &$22.8$   &$<38.2$        &$1.295$       &$1.055$       &$4.3^{+4.3}_{-2.8}\pm0.2$                     &$<10.8$\\
  $4.575$     &$47.7$      &$2$         &$2$    &$1$    &$22.6$   &$<22.5$        &$1.314$       &$1.055$       &$4.3^{+7.7}_{-3.8}\pm0.2$                     &$<14.5$\\
  $4.600$     &$570.0$     &$5$         &$34$   &$19$   &$22.4$   &$<23.0$        &$1.323$       &$1.055$       &$-0.8^{+0.9}_{-0.6}\pm0.1$                    &$<1.2$\\
  \hline
  \hline
    \end{tabular}
    \label{ResultsEtaUpper}
\end{table*}

The Born cross section is calculated by:
\begin{equation}
\sigma^{B}=\frac{N^\mathrm{obs}}{\mathcal{L}_\mathrm{int}\cdot (1+\delta^{r})\cdot (1+\delta^{v})
\cdot \epsilon \cdot \cal B } ,
\label{eqcross}
\end{equation}
where $N^\mathrm{obs}$ is the number of observed signal events, $\mathcal{L}_\mathrm{int}$ is the integrated luminosity,
$(1+\delta^{r})$ is the ISR correction factor which is obtained by QED calculation~\cite{QED}
and taking the line shape of the Born cross section measured by the Belle experiment~\cite{belle}.
The vacuum polarization (VP) factor $(1+\delta^{v})$ is taken from a QED calculation with an accuracy of $0.5\%$~\cite{VP},
$\epsilon$ is the detection efficiency including reconstruction and all selection criteria,
$\cal B$ is the product branching ratio, and $\BR(J/\psi \to \LL)\cdot \BR(\eta \to \gamma\gamma)$,
taken from the Particle Data Group (PDG)~\cite{PDG}.

The final Born cross sections of $e^+e^-\to \eta J/\psi$ at energy points
with a statistically significant observation of signal events are listed in Table~\ref{ResultsEta}.

For the other energy points where the $\eta$ signal is not significant, we set
upper limits at the 90\% C.L. on the Born cross section.
The upper limit is calculated by a frequentist method with a profile likelihood treatment
of systematic uncertainties taken into account in the efficiency uncertainty,
which is implemented by a C++ class TROLKE in the ROOT framework~\cite{TROLKE}.
The numbers of observed signal events and estimated background events are assumed to follow a Poisson distribution,
and the efficiencies are assumed have Gaussian uncertainties.
Since the number of background events can be estimated from either the $\eta$ or $J/\psi$ sideband events,
the one with the larger upper limit on the Born cross section is taken as the final result as a conservative estimation.
The results on the upper limits are listed in Table ~\ref{ResultsEtaUpper}.

Since there is no significant signal of $e^{+}e^{-} \to \pi^{0}J/\psi$ observed at any energy,
we set upper limits at the 90\% C.L. on the Born cross section.
The number of observed events is obtained by counting the entries in the $\pi^0$ signal
region ($0.120<M(\gamma\gamma)<0.150$ GeV/c$^2$).
The number of background events in the signal region is estimated by counting the number
of events in the $\pi^0$ sideband regions ($0.055<M(\gamma\gamma)<0.115$ GeV/c$^2$ and
$0.155<M(\gamma\gamma)<0.215$ GeV/c$^2$) or $J/\psi$ sideband regions
(with an additional $\pi^0$ signal mass window requirement).
The same frequentist method is implemented to extract the upper limits.
The results and the related variables used to calculate the upper limit are listed in Table~\ref{Npi0Jpsi}.

\begin{table*}[htbp]
  \centering
  \caption{Upper limits of $e^{+}e^{-} \to \pi^{0} J/\psi$. The table shows the number of observed events in the $\pi^{0}$ signal region $N^\mathrm{sg}$, number of events in $\pi^{0}$ sideband $N^\mathrm{sb}_{\pi^{0}}$, and in $J/\psi$ sideband $N^\mathrm{sb}_{J/\psi}$, efficiency $\epsilon$, the upper limit of signal events with the consideration of the selection efficiency $N^\mathrm{up}(\mu^{+}\mu^{-})/\epsilon$ (at the $90\%$ C.L.) and the upper limit of Born cross sections $\sigma^{B}_\mathrm{up}$ (at the $90\%$ C.L.).}
  \begin{tabular}{ccccccccc}
  \hline
  \hline
  \mystrut
  ~$\sqrt{s}$(GeV)~ &~$N^\mathrm{sg}$~  &~$N^\mathrm{sb}_{\pi^{0}}$~  &~$N^\mathrm{sb}_{J/\psi}$~ &~$\epsilon(\%)$~   &~~~$N^\mathrm{up}/\epsilon$~  &~$(1+\delta^{r})$~ &~$(1+\delta^{v})$~    &~$\sigma^{B}_\mathrm{up}$(pb)~\\
  \hline
  $3.810$         &$1$       &$4$                 &$1$               &$16.9$            &$<20.2$                      &$1.243$             &$1.056$              &$<5.2$\\
  $3.900$         &$0$       &$1$                 &$2$               &$29.2$            &$<6.0$                       &$0.775$             &$1.049$              &$<2.4$\\
  $4.090$         &$0$       &$0$                 &$2$               &$25.7$            &$<7.8$                       &$1.078$             &$1.052$              &$<2.2$\\
  $4.190$         &$0$       &$0$                 &$0$               &$29.9$            &$<6.7$                       &$0.866$             &$1.056$              &$<2.9$\\
  $4.210$         &$1$       &$1$                 &$1$               &$29.0$            &$<11.8$                      &$0.914$             &$1.057$              &$<3.8$\\
  $4.220$         &$0$       &$1$                 &$0$               &$28.5$            &$<7.0$                       &$0.937$             &$1.057$              &$<2.2$\\
  $4.230$         &$4$       &$16$                &$13$              &$28.1$            &$<18.5$                      &$0.960$             &$1.056$              &$<0.3$\\
  $4.245$         &$1$       &$1$                 &$2$               &$27.3$            &$<12.6$                      &$0.992$             &$1.056$              &$<3.7$\\
  $4.260$         &$3$       &$8$                 &$10$              &$26.5$            &$<18.8$                      &$1.021$             &$1.054$              &$<0.4$\\
  $4.310$         &$0$       &$0$                 &$0$               &$24.6$            &$<8.3$                       &$1.105$             &$1.053$              &$<2.7$\\
  $4.360$         &$2$       &$3$                 &$4$               &$23.5$            &$<19.9$                      &$1.168$             &$1.051$              &$<0.5$\\
  $4.390$         &$1$       &$0$                 &$1$               &$23.1$            &$<16.0$                      &$1.198$             &$1.051$              &$<3.9$\\
  $4.420$         &$2$       &$7$                 &$20$              &$22.7$            &$<16.3$                      &$1.225$             &$1.053$              &$<0.2$\\
  $4.470$         &$0$       &$0$                 &$4$               &$22.3$            &$<8.9$                       &$1.258$             &$1.055$              &$<1.0$\\
  $4.530$         &$0$       &$1$                 &$2$               &$21.8$            &$<8.9$                       &$1.295$             &$1.055$              &$<0.9$\\
  $4.575$         &$0$       &$0$                 &$2$               &$21.7$            &$<9.2$                       &$1.314$             &$1.055$              &$<2.4$\\
  $4.600$         &$3$       &$5$                 &$7$               &$21.6$            &$<26.2$                      &$1.323$             &$1.055$              &$<0.6$\\
  \hline
  \hline
    \end{tabular}
    \label{Npi0Jpsi}
\end{table*}

\section{Systematic Uncertainties}
Several sources of systematic uncertainties are considered in the measurement of the Born cross sections.
These include differences between data and MC simulation for the tracking efficiency, photon detection,
kinematic fit, mass window requirement, the fit procedure, the shower depth in the MUC, MC simulation of
the ISR correction factor and vacuum polarization factor, as well as uncertainties in the branching fractions of
intermediate state decays and in the luminosity measurements.

(a) \emph{Tracking:} The uncertainty of the tracking efficiency is investigated using a control sample
$\psi(3686)\to\pi^{+}\pi^{-} J/\psi$ with the subsequent decay of $J/\psi\to\ell^+\ell^-$.
The difference in tracking efficiency for the lepton reconstruction between data and MC simulation is estimated to be 1\% per track.
So, 2\% is taken as the systematic uncertainty for the two leptons.

(b) \emph{Photon detection efficiency:} The uncertainty due to the photon detection and reconstruction efficiency is 1\% per photon~\cite{Gamma}.
This value is determined from studies using background-free control samples $J/\psi\to\rho^{0}\pi^{0}$ and $e^{+}e^{-} \to \gamma\gamma$.
Therefore, an uncertainty of 2\% is taken for the detection efficiency of two photons.

(c) \emph{Kinematic fit:} In order to reduce the difference on the $4C$ kinematic fit $\chi^{2}_{4C}$ between data and MC simulations, the track helix
parameters ($\phi_{0}$, $\kappa$, $\tan \lambda$) of simulated tracks have been corrected, where $\phi_{0}$ is the azimuthal angle that specifies the pivot with
respect to the helix center, $\kappa$ is the reciprocal of the transverse momentum, and $\tan\lambda$ is the slope of the track.
The correction factors are obtained from a nearly background-free sample of
$e^{+}e^{-} \to \pi^{+}\pi^{-}J/\psi$ and $J/\psi \to e^{+}e^{-}/\mu^{+}\mu^{-}$ at $\sqrt{s}$ = 4.230 GeV.
An alternative detector efficiency is evaluated with the same MC samples, but without helix parameters corrections.
The difference in this efficiency from its nominal value is taken to be the uncertainty due to the $4C$ kinematic fit requirement~\cite{Helix}.

(d) \emph{Mass window requirements:} A mass window requirement on the $\ell^+\ell^-$ invariant mass introduces a systematic uncertainty on its efficiency.
The $J/\psi$ signal at $\sqrt{s}$ = 4.230 GeV is fitted with a MC shape convoluted with a Gaussian function, where the
parameters of the Gaussian function are left free in the fit.
To evaluate the systematic effects on the mass window requirement, the invariant mass of $\ell^+\ell^-$
in MC samples are smeared with a Gaussian function, where the parameters of the Gaussian function are
obtained from the fit.
The difference in the efficiencies between the signal MC sample with and without mass
resolution smearing is 0.2\% in the $\mu^{+}\mu^{-}$ mode and 0.1\% in the $e^{+}e^{-}$ mode,
and is taken as the systematic uncertainty.

\begin{table*}[htbp]
  \centering
  \caption{Summary of systematic uncertainties ($\%$) in the cross section of $e^{+}e^{-} \to \eta J/\psi$
  for energies with significant signal in the $\mu^{+}\mu^{-}$($e^{+}e^{-}$) mode. The common uncertainties (Luminosity, Tracking, Photon, Branching fraction and Others) between the two modes are shown together.}
  \begin{tabular}{ccccccccc}
  \hline
  \hline
  Source/$\sqrt{s}$(GeV)    ~~~~&~~~~4.190   ~~~~&~~~~4.210   ~~~~&~~4.220   ~~~~&~~~~4.230   ~~~~&~~~~4.245    ~~~~&~~~~4.260    ~~~~&~~~~4.360    ~~~~&~~~~4.420~~~~    \\
  \hline
  Luminosity                &1.0          &1.0          &1.0          &1.0          &1.0           &1.0           &1.0          &1.0\\
  Tracking                  &2.0          &2.0          &2.0          &2.0          &2.0           &2.0           &2.0          &2.0\\
  Photon                    &2.0          &2.0          &2.0          &2.0          &2.0           &2.0           &2.0          &2.0\\
  Kinematic fit             &0.4 (0.4)    &0.4 (0.4)    &0.4 (0.3)    &0.4 (0.3)    &0.5 (0.5)     &0.4 (0.4)     &0.3 (0.4)    &0.4 (0.4)\\
  Mass window               &0.2 (0.1)    &0.2 (0.1)    &0.2 (0.1)    &0.2 (0.1)    &0.2 (0.1)     &0.2 (0.1)     &0.2 (0.1)    &0.2 (0.1)\\
  Fitting range             &0.0 (1.0)    &0.4 (0.7)    &0.3 (2.6)    &0.1 (2.2)    &0.0 (1.0)     &0.0 (0.6)     &8.6 (7.5)    &0.7 (2.1)\\
  Signal shape              &0.3 (1.1)    &0.3 (1.1)    &0.3 (1.1)    &0.3 (1.1)    &0.3 (1.1)     &0.3 (1.1)     &0.3 (1.1)    &0.3 (1.1)\\
  Background shape          &4.6 (0.1)    &3.9 (6.8)    &2.8 (0.0)    &0.0 (0.1)    &9.7 (9.3)     &0.2 (0.0)     &0.5 (0.4)    &0.1 (0.2)\\
  ISR factor                &0.6 (0.3)    &4.3 (4.2)    &4.7 (6.0)    &4.2 (5.9)    &4.0 (3.6)     &6.6 (5.8)     &9.4 (9.1)    &10.5 (7.7)\\
  Branching fraction        &0.8          &0.8          &0.8          &0.8          &0.8           &0.8           &0.8          &0.8\\
  Others                    &1.0          &1.0          &1.0          &1.0          &1.0           &1.0           &1.0          &1.0\\
  Sum                       &5.7 (3.6)    &6.7 (8.7)    &6.4 (7.4)    &5.3 (7.2)    &11.0 (10.6)   &7.4 (6.8)     &13.2 (12.3)  &11.0 (8.7)\\
  \hline
  \hline
    \end{tabular}
    \label{systematic1}
\end{table*}

\begin{table*}[htbp]
  \centering
  \caption{Summary of systematic uncertainties ($\%$) in the upper limit on cross section of $e^{+}e^{-} \to \eta J/\psi$ in $\mu^{+}\mu^{-}$ mode.}
  \begin{tabular}{cccccccccc}
  \hline
  \hline
  Source/$\sqrt{s}$(GeV)    ~~&~~3.810    ~~&~~3.900    ~~&~~4.090 ~~&~~4.310    ~~&~~4.390   ~~&~~4.470  ~~&~~4.530  ~~&~~4.575  ~~&~~4.600\\
  \hline
  Luminosity                &1.0          &1.0          &1.0       &1.0          &1.0         &1.0        &1.0        &1.0        &1.0       \\
  Tracking                  &2.0          &2.0          &2.0       &2.0          &2.0         &2.0        &2.0        &2.0        &2.0       \\
  Photon                    &2.0          &2.0          &2.0       &2.0          &2.0         &2.0        &2.0        &2.0        &2.0       \\
  Kinematic fit             &0.4          &0.4          &0.4       &0.4          &0.4         &0.4        &0.4        &0.4        &0.4       \\
  Mass window               &0.2          &0.2          &0.2       &0.2          &0.2         &0.2        &0.2        &0.2        &0.2       \\
  ISR factor                &0.2          &1.3          &6.3       &0.4          &9.0         &2.8        &1.0        &1.2        &0.8    \\
  Branching fraction        &0.8          &0.8          &0.8       &0.8          &0.8         &0.8        &0.8        &0.8        &0.8       \\
  Others                    &1.0          &1.0          &1.0       &1.0          &1.0         &1.0        &1.0        &1.0        &1.0       \\
  Sum                       &3.3          &3.5          &7.1       &3.3          &9.6         &4.3        &3.4        &3.5        &3.4    \\
  \hline
  \hline
    \end{tabular}
    \label{systematic2}
\end{table*}

\begin{table*}[htbp]
  \centering
  \caption{Summary of systematic uncertainties ($\%$) in the cross section of $e^{+}e^{-} \to \pi^{0} J/\psi$.}
  \begin{tabular}{cccccccccccccccccc}
  \hline
  \hline
  Source/$\sqrt{s}$(GeV)  &3.810  &3.900  &4.090  &4.190  &4.210  &4.220  &4.230  &4.245  &4.260  &4.310  &4.360  &4.390  &4.420  &4.470 &4.530 &4.575 &4.600\\
  \hline
  Luminosity              &1.0    &1.0    &1.0    &1.0    &1.0    &1.0    &1.0    &1.0    &1.0    &1.0    &1.0    &1.0    &1.0    &1.0   &1.0   &1.0      &1.0      \\
  Tracking                &2.0    &2.0    &2.0    &2.0    &2.0    &2.0    &2.0    &2.0    &2.0    &2.0    &2.0    &2.0    &2.0    &2.0   &2.0   &2.0      &2.0      \\
  Photon                  &2.0    &2.0    &2.0    &2.0    &2.0    &2.0    &2.0    &2.0    &2.0    &2.0    &2.0    &2.0    &2.0    &2.0   &2.0   &2.0      &2.0      \\
  Kinematic fit           &0.4    &0.4    &0.4    &0.4    &0.4    &0.4    &0.4    &0.4    &0.4    &0.4    &0.4    &0.4    &0.4    &0.4   &0.4   &0.4      &0.4      \\
  Mass window             &0.2    &0.2    &0.2    &0.2    &0.2    &0.2    &0.2    &0.2    &0.2    &0.2    &0.2    &0.2    &0.2    &0.2   &0.2   &0.2      &0.2      \\
  MUC cut                 &1.2    &1.2    &1.2    &1.2    &1.2    &1.2    &1.2    &1.2    &1.2    &1.2    &1.2    &1.2    &1.2    &1.2   &1.2   &1.2      &1.2      \\
  ISR factor              &0.2    &1.1    &6.5    &0.3    &4.6    &5.7    &3.9    &4.1    &6.7    &0.8    &9.6    &8.7    &7.9    &1.0   &0.7   &0.5   &0.7   \\
  Branching fraction      &0.6    &0.6    &0.6    &0.6    &0.6    &0.6    &0.6    &0.6    &0.6    &0.6    &0.6    &0.6    &0.6    &0.6   &0.6   &0.6      &0.6      \\
  Others                  &0.6    &0.6    &0.6    &0.6    &0.6    &0.6    &0.6    &0.6    &0.6    &0.6    &0.6    &0.6    &0.6    &0.6   &0.6   &0.6      &0.6      \\
  Sum                     &3.5    &3.6    &7.4    &3.5    &5.8    &6.7    &5.2    &5.4    &7.5    &3.6    &10.2   &9.4    &8.6    &3.6   &3.5   &3.5   &3.5   \\
  \hline
  \hline
    \end{tabular}
    \label{systematic3}
\end{table*}

(e) \emph{Fitting procedure:} For the eight data samples with clearly observed $\eta$ signals, fits to the two photon
invariant mass $M(\gamma\gamma)$ are performed to extract the number of $e^+e^-\to \eta J/\psi$ decays.
The following three aspects are considered when evaluating the systematic uncertainty associated with
the fit procedure.
(1) {\it Fitting range:} In the fit, the $M(\gamma\gamma)$ is fitted in a region from 0.2 to 0.9 GeV/c$^2$.
  An alternative fit with a different fit range, from 0.25 to 0.85 GeV/c$^2$, is performed. The differences
  in the yield are treated as the systematic uncertainty from the fit range.
(2) {\it Signal shape:} In the fit, the signal shape is described by a shape obtained from a MC simulation convoluted
   with a Gaussian function. An alternative fit with a Crystal Ball function~\cite{CrystalBall} for the $\eta$ signal shape
   is performed, where the parameters of the Crystal Ball function at different CM energies
   are fixed to those obtained from the fit of the $\eta$ signal at $\sqrt{s}$ = 4.230 GeV.
   The difference in the yield with respect to the nominal fit is considered as the systematic uncertainty from the signal shape.
(3) {\it Background shape:} In the fit, background shapes are described as a second-order
   polynomial function. The fit with a third-order polynomial function for the background shape
   is used to estimate its uncertainty.
For the data sets where no evident $\eta$ signal is found,
the frequentist method is employed to determine upper limits on the Born cross section,
and the numbers of signal and background events are obtained by counting the entries in signal and sideband regions.
Two different sideband regions, either the $\eta$ sideband region or
the $J/\psi$ sideband region, are used to estimate the uncertainty from the background shape.
The systematic uncertainty associated with the background shape has been considered by taking the most conservative upper limit as the final result.

(f) \emph{MUC requirement:} In the search for the process $e^{+}e^{-} \to \pi^{0} J/\psi$,
an additional requirement on the hit depth in the MUC for muon tracks was imposed to remove
the background from $e^{+}e^{-} \to \pi^+\pi^-\pi^{0}$.  By studying the control sample of
$e^{+}e^{-} \to \pi^{+}\pi^{-} J/\psi$ with a subsequent decay of $J/\psi\to \mu^{+}\mu^{-}$ at $\sqrt{s}$ = 4.230 GeV,
the efficiency difference of this requirement between data and MC sample was found to be (9.0 $\pm$ 1.2)$\%$.
The MC efficiency has been corrected for this difference and a value of 1.2\% is taken as the corresponding systematic uncertainty.

(g) \emph{ISR factor:} The uncertainties of the line shape of the cross section used in the KKMC generator
introduce uncertainties in both the radiative correction factor and the efficiency.
In the nominal results, the line shape of the cross section is taken from the fit result from the Belle experiment~\cite{belle}.
We have also performed a new fit with three incoherent Breit-Wigner functions, including the $Y(4360)$
and a second order polynomial function, to the same observed cross section $\sigma(e^{+}e^{-}\to \eta J/\psi)$, where the parameters of the Breit-Wigner
functions are left free in the fit.
With this line shape of the cross section, the variations in $(1+\delta^{r})\times\epsilon$ are taken as the uncertainties.

(h) \emph{Luminosity:} The integrated luminosity of data samples used in this analysis are measured using large angle
Bhabha events, and the corresponding uncertainties are estimated to be 1.0$\%$~\cite{Luminosity}.

(i) \emph{Branching fractions:} The experimental uncertainties in the branching fractions for the processes $J/\psi \to \LL$, $\eta \to \gamma \gamma$
and $\pi^{0} \to \gamma \gamma$ are taken from the PDG~\cite{PDG}.

(j) \emph{Other systematic uncertainties:} Other sources of systematic uncertainties include the trigger efficiency,
event start time determination and final-state-radiation simulation.
The total systematic uncertainty due to these sources is estimated to be less than $1.0\%$.

Assuming all of the above systematic uncertainties, shown in Table~\ref{systematic1}, Table~\ref{systematic2} and Table~\ref{systematic3}, are
independent, the total systematic uncertainties are obtained by adding the individual
uncertainties in quadrature.

For the energy points where statistically significant signal yields were found, the results from the two $J/\psi$ decay modes
are found to be consistent. The combined cross sections are calculated by considering the correlation of uncertainties between these two measurements~\cite{Combined}
and the results are also listed in Table~\ref{ResultsEta}.

\section{Summary and Discussion}
In summary, using data samples collected with the BESIII detector at energies from 3.810 to 4.600 GeV,
we performed an analysis of $e^{+}e^{-} \to \eta J/\psi$.
Statistically significant $\eta$ signals are observed at $\sqrt{s}$ = 4.190, 4210, 4220, 4230, 4245, 4260, 4360 and 4420 GeV,
and the corresponding Born cross sections are measured.
In addition, we searched for the process $e^{+}e^{-} \to \pi^0 J/\psi$. No significant signals are
observed and the upper limits at the 90\% C.L. on the Born cross section are set.

\begin{figure}[htbp]
\begin{center}
\begin{overpic}[width=7.0cm,height=5.0cm,angle=0]{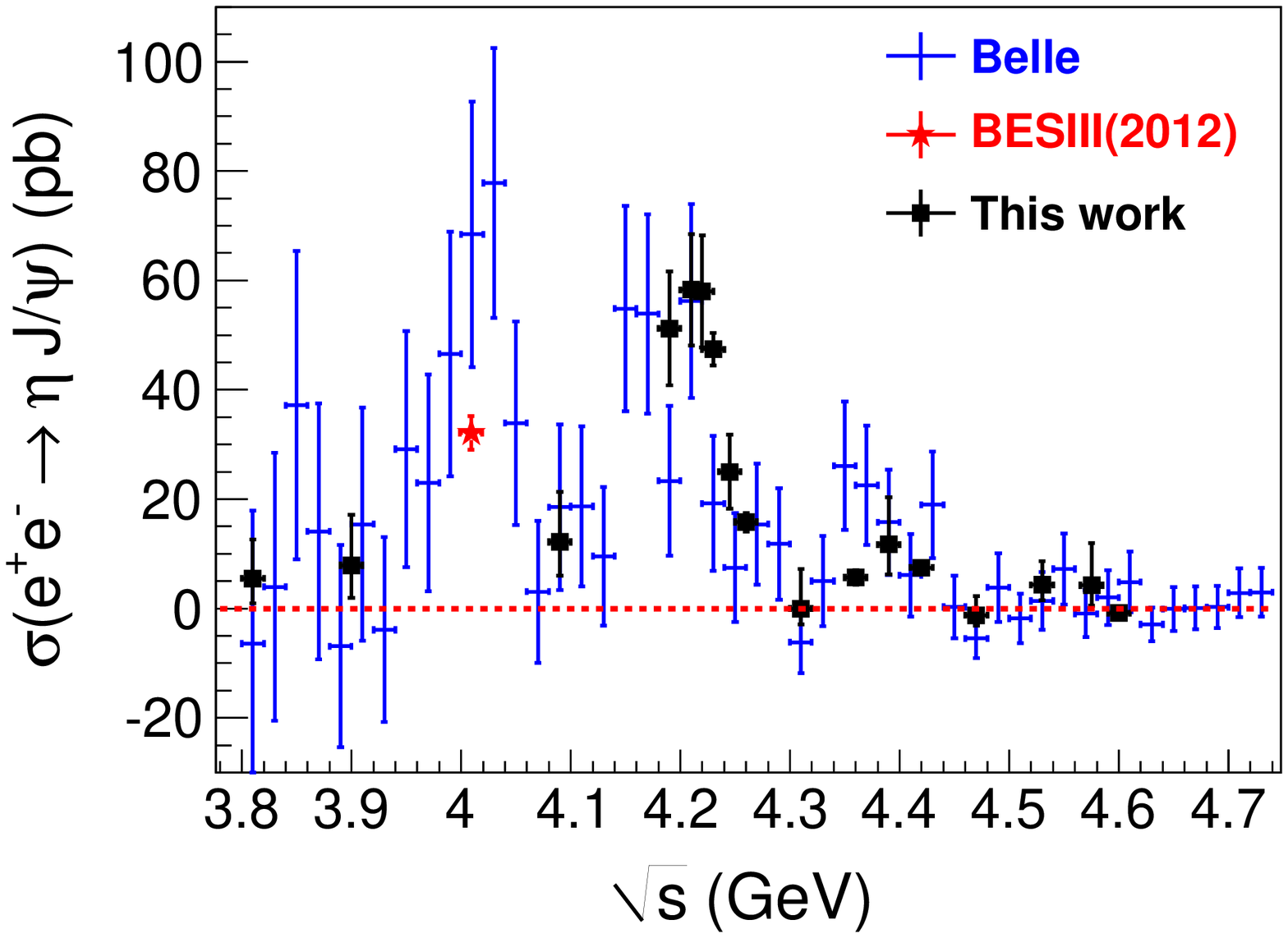}
\put(19,57){\large\bf (a)}
\end{overpic}
\begin{overpic}[width=7.0cm,height=5.0cm,angle=0]{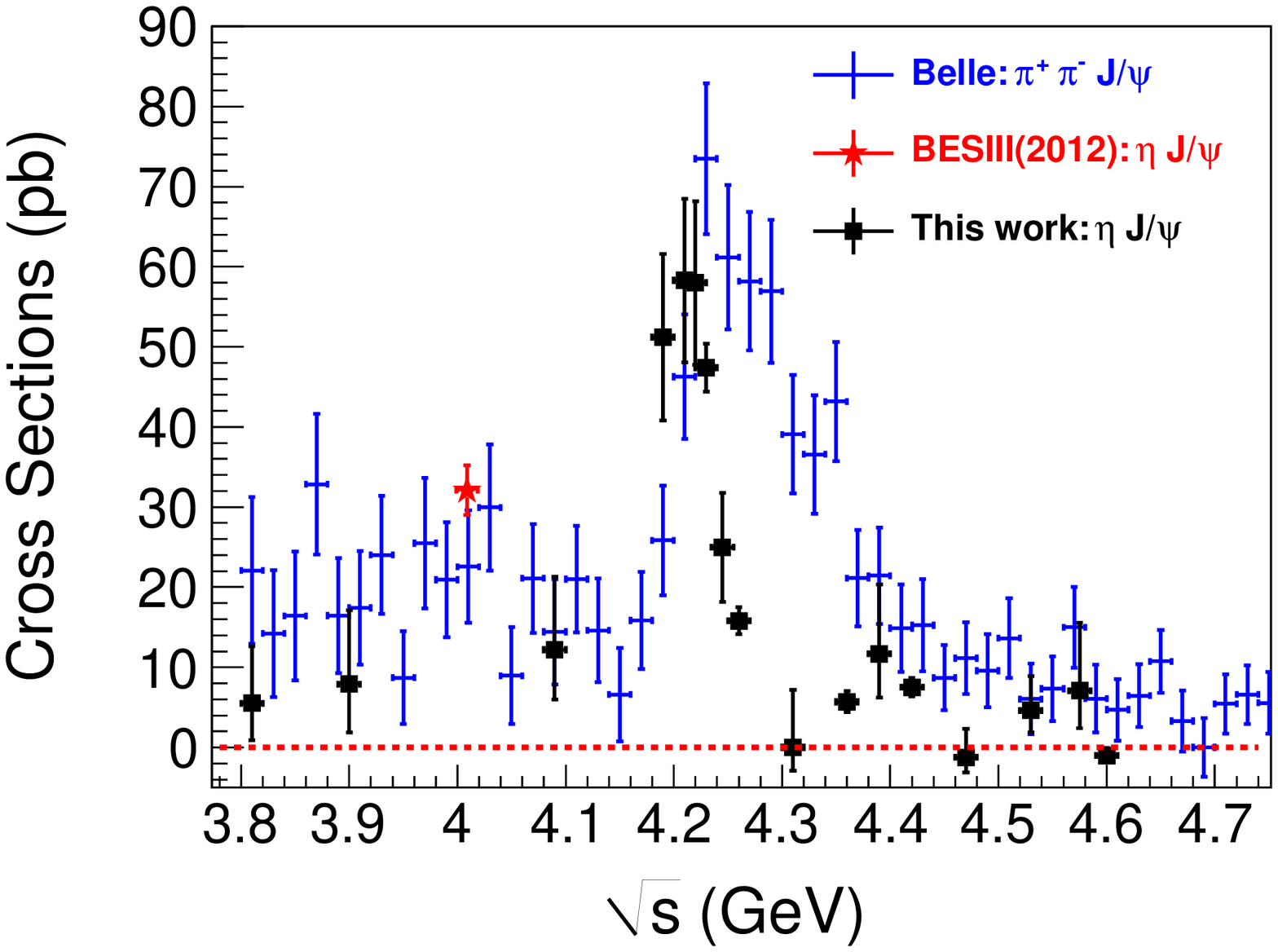}
\put(19,57){\large\bf (b)}
\end{overpic}
\end{center}
\caption{A comparison of the measured Born cross section of $e^{+}e^{-} \to \eta J/\psi$ to (a) that of a previous measurement~\cite{4040,belle}, (b) that of $e^{+}e^{-}\to\pi^{+}\pi^{-} J/\psi$ from Belle~\cite{Y(4260)21}. In these two plots, the black square dots are the results of $\eta J/\psi$ obtained in this work and the red star dots are from BESIII(2012). The blue dots are results of $\eta J/\psi$ (a) and $\pi^{+}\pi^{-} J/\psi$ (b) from Belle. The errors are statistical only for Belle's results, and are final combined uncertainties for BESIII's results.}\label{BES_BELLE}
\end{figure}

A comparison of the Born cross sections $\sigma(e^{+}e^{-} \to \eta J/\psi)$ in this measurement to that of previous results~\cite{4040,belle} is shown in Fig.~\ref{BES_BELLE} (a),
and a very good agreement is achieved.
The measured Born cross sections are also compared to that of $e^{+}e^{-} \to \pi^{+}\pi^{-} J/\psi$ obtained from Belle~\cite{Y(4260)21} as shown in Fig.~\ref{BES_BELLE} (b).
Different lineshapes are observed in these two processes, which indicate that the production mechanism of the $\eta J/\psi$
clearly differs from that of $\pi^{+}\pi^{-}J/\psi$ in the vicinity of $\sqrt{s}$ = 4.1-4.6 GeV.
This could indicate the existence of a rich spectrum of $Y$ states in this energy region with different coupling strengths to the various decay modes.

The ratio of the Born cross section at 4.260 GeV to that at 4.230 GeV,
$R^{4.260}_{4.230}(e^+e^-\to\eta J/\psi)=\frac{\sigma^{4.260}(e^+e^-\to\eta J/\psi)}{\sigma^{4.230}(e^+e^-\to\eta J/\psi)}$,
is calculated to be $0.33 \pm 0.04$ (common systematic uncertainties cancel in the calculation),
which is found to agree very well with the ratio,
$R^{4.260}_{4.230}(e^+e^-\to\omega \chi_{c0})=\frac{\sigma^{4.260}(e^+e^-\to\omega \chi_{c0})}{\sigma^{4.230}(e^+e^-\to\omega \chi_{c0})}$ = $0.43~\pm~0.13$,
of the process $e^+e^-\to\omega \chi_{c0}$~\cite{OmegaChicj}. This may indicate that the production
of $\eta J/\psi$ and $\omega \chi_{c0}$ are from the same source. More data around this energy region
may be useful to clarify this interpretation. Compared with a theoretical prediction~\cite{Theory2} that considers open charm effects on the exclusive cross section line shapes of $e^{+}e^{-}\to \eta J/\psi$ and $\pi^{0} J/\psi$,
our results on $\eta J/\psi$ are within the range of the theoretical prediction,
and the obtained $\pi^{0} J/\psi$ upper limits are higher by a factor of 50 than that of the theoretical prediction.
More data samples will be helpful to test the predicted cross section of $e^{+}e^{-}\to \pi^{0}J/\psi$.

\acknowledgments
The BESIII collaboration thanks the staff of BEPCII and the IHEP
computing center for their strong support. This work is supported in
part by National Key Basic Research Program of China under Contract
No.~2015CB856700; National Natural Science Foundation of China (NSFC)
under Contracts Nos.~11125525, 11235011, 11322544, 11335008, 11425524;
the Chinese Academy of Sciences (CAS) Large-Scale Scientific Facility
Program; Joint Large-Scale Scientific Facility Funds of the NSFC and
CAS under Contracts Nos.~11179007, U1232201, U1332201; CAS under
Contracts Nos.~KJCX2-YW-N29, KJCX2-YW-N45; 100 Talents Program of CAS;
INPAC and Shanghai Key Laboratory for Particle Physics and Cosmology;
German Research Foundation DFG under Contract No.~Collaborative
Research Center CRC-1044; Istituto Nazionale di Fisica Nucleare,
Italy; Ministry of Development of Turkey under Contract
No.~DPT2006K-120470; Russian Foundation for Basic Research under
Contract No. 14-07-91152; U.S.~Department of Energy under Contracts
Nos.~DE-FG02-04ER41291, DE-FG02-05ER41374, DE-FG02-94ER40823,
DESC0010118; U.S.~National Science Foundation; University of Groningen
(RuG) and the Helmholtzzentrum fuer Schwerionenforschung GmbH (GSI),
Darmstadt; WCU Program of National Research Foundation of Korea under
Contract No.~R32-2008-000-10155-0.

\end{document}